\documentclass[a4paper,12pt]{llncs}
\usepackage{llncsdoc}
\usepackage{makeidx}
\usepackage{amsmath,amssymb}
\usepackage{hyperref}
\usepackage[left=2.5cm,top=2.5cm,right=2.5cm,bottom=2.5cm]{geometry}
\usepackage{epsfig}
\usepackage{color}
\usepackage{nicefrac}
\usepackage{multirow}
\usepackage{hyperref}
\usepackage{url}
\usepackage{tikz}
\usetikzlibrary{shapes,arrows,shadows,positioning,matrix,backgrounds}
\usepackage[caption=false]{subfig}

\newcommand{\Erdos}{Erd\H{o}s }
\newcommand{\Renyi}{R\'enyi }

\newcommand{\vecw}{{\bf w}}

\date{}
\begin{document}
\title{\Large{Mining and Modeling Character Networks}\thanks{Research supported by grants from NSERC and Ryerson University; Gleich and
Hou's work were supported by NSF CAREER Award CCF-1149756, IIS-1546488, CCF-093937, and DARPA SIMPLEX.}}
\author{Anthony Bonato \inst{1}\and David Ryan D'Angelo \inst{1}\and \\ Ethan R.\ Elenberg \inst{2} \and David F.~Gleich \inst{3} \and Yangyang Hou  \inst{3}}
\institute{Ryerson University \and University of Texas at Austin \and Purdue University}
\maketitle
\begin{abstract}
We investigate social networks of characters found in cultural works such as novels and films. These \emph{character networks} exhibit many of the properties of complex networks such as skewed
degree distribution and community structure, but may be of relatively small order with a high multiplicity of edges. Building on recent work of Beveridge and Shan \cite{beveridge}, we consider graph extraction,
visualization, and network statistics for three novels: \emph{Twilight} by Stephanie Meyer, Steven King's \emph{The Stand,} and J.K. Rowling's \emph{Harry Potter and the Goblet of Fire.} Coupling
with 800 character networks from films found in the \url{http://moviegalaxies.com/} database, we compare the data sets to simulations from various stochastic complex networks models including
random graphs with given expected degrees (also known as the Chung-Lu model), the configuration model, and the preferential attachment model. Using machine learning techniques based on motif (or
small subgraph) counts, we determine that the Chung-Lu model best fits character networks and we conjecture why this may be the case.
\end{abstract}

\section{Introduction}\label{intro}

Complex networks lie at the intersection of several disciplines and have found broad application within the study of social networks. In social networks, nodes represents agents, and edges correspond
to some kind of social interaction such as friendship or following. For more on complex networks and on-line social networks, the reader is directed to the book \cite{bon1} and the survey
\cite{bon2}.

In the present paper, we consider social networks arising in the context of cultural works such as novels or movies. In these \emph{character networks}, nodes represent characters in a specified
fictional or non-fictional work such as a novel, script, biography, or story, with edges between characters determined by their interaction within the work. We also consider character networks as
weighted graphs, where the weights are positive integers specifying the co-appearance or co-occurrence of character names within a specified range of the text or scenes (such as being within fifteen
words of each other; see \cite{beveridge}). Not surprisingly, character networks are typically of smaller order than many other types of complex networks. Nevertheless, they still exhibit many of the
interesting features of complex networks including clearly defined community structure, with communities centered on the various protagonists of the story, skewed degree distributions, focused on the
most important characters, and dynamics. Character networks defined over larger fictional universes, such as the Marvel Universe, even grow to over 10,000 nodes~\cite{alberich,gleiser}.

There is an emerging approach using the tools of graph theory and big data to mine and model character networks. This new topic reflects the ease of access of cultural works in electronic formats,
and the efficacy of big data-theoretic algorithms. Our approach in this work is study new networks with these tools to replicate some of the findings as well as study network models of these data.

First, we wish to study the complexity of these character networks through graph mining. Our approach here is more a microscopic view of an individual work's network. We focus on three well known
novels: \emph{Twilight} by Stephanie Meyer, Steven King's \emph{The Stand}, and J.K. Rowling's \emph{Harry Potter and the Goblet of Fire}. Various complex network statistics, such as diameter and
clustering coefficient, are presented along with centrality metrics (such as PageRank and betweenness, paralleling the approach of \cite{beveridge}) that predict the major characters within each
book. See Section~\ref{exp} for the methodology used, and Section~\ref{results} has a summary of our results.

The second part of our approach is to compare and contrast the character networks with several well known stochastic network models. Hence, in this approach, we take a broader, macroscopic view of
the structure of a larger sample of character networks. Using motifs (that is, small subgraph counts), eigenvalues, and machine learning techniques, we develop an approach for model selection for
character networks. The models considered were the configuration model, preferential attachment model, the Chung-Lu model for random graphs with given expected degree sequences, and the binomial
random graph (as a control). The parameters of the models were chosen as to equal the number of nodes and average degree of the character network data sets. Model selection was conducted for the
three novels described above, and also for a set of 800 networks arising from movies in the \url{http://moviegalaxies.com/} database \cite{movieGalaxies}. Our results show consistent selection of the Chung-Lu model as
the most realistic, with a clear separation between the models. We will discuss possible interpretations and implications of our results in the final section.

We consider undirected graphs throughout the paper. For background on graph theory, the reader is directed to~\cite{west}. Additional background on machine learning can be found in
\cite{statisticalLearning,sd}.

\subsection{Previous Work}\label{lit}

Quantitative methods have now emerged as a modern tool for literary analysis.  Literary theories are now supported, debated, and refuted based on data~\cite{brit}.  In recent work, Reagan et
al.~\cite{shapes} implement data mining techniques inspired by Kurt Vonnegut's theory of the shape of stories. Vonnegut suggested graphing fictional works based on the fortune of the main character's
experiences over the passage of time in the story. Using text sentiment analysis, Reagan et al.\ scored the emotional content over the course of a novel based on the occurrence of select words in the
labMT data set for 1,737 books from the Project Gutenberg database. They found the majority of emotional arcs resided in six classes. 
In a study of 60 novels, including Jane
Austin's \textit{Pride and Prejudice}, Dames et al.~\cite{brit} determined that the type of narrative is a good predictor for social network structure among characters.

In \cite{beveridge}, Beveridge and Shan applied network algorithms on the social network they generated from \textit{A Storm of Swords}, the third novel in George R.R.\ Martin's \textit{A Song of Ice
and Fire} series (which is the literary origin of the HBO drama \emph{Game of Thrones}). Metrics such as PageRank, closeness, betweenness centrality, and modularity provided an empirical approach to
determine communities and key characters within the network. Work done by Ribeiro et al.~\cite{rings} focuses on examining structural properties, such as assortivity and transitivity, of communities
in the social network of J.R.R.\ Tolkien's \emph{The Lord of the Rings} (which included that unabridged novel, along with text from \emph{The Hobbit} and \emph{The Silmarillion}). Beyond static
networks, Agarwal et al.\ \cite{alice} analyze the dynamic network for \textit{Alice in Wonderland}, defined by the mining of the ten chapters independently of each other. 
Such analysis may be
important in determining characters with low global importance metrics who are significantly important for part of the story. Deviating from the extraction of
character networks, Sack \cite{sack} provides a social network generation model for narratives through the concept of \emph{structural balance theory} using signed edges between characters.

\section{Experimental Design and Methods}\label{exp}

The twin goals of our experiments are to highlight some of the complexities present in character networks via their network properties and to determine a possible synthetic model of the character networks.

\subsection{Network Properties}

We use the Gephi open source software package to extract communities and compute various network statistics from character networks.  These analyses are all done on weighted, undirected, graphs. For
community analysis, we use modularity and the Louvain method.
Centrality measures are a classic tool in social network analysis to determine the important individuals. They have been found to also serve the same role in character networks.
We consider weighted degree, closeness, betweenness, eigencentrality, and PageRank centrality. We briefly review these methods; see \cite{bon1} for more background on complex network properties.
The \emph{closeness} of a node $u$ is the average distance between $u$ and all other nodes (here distance is the standard shortest path metric in graph theory). The \emph{betweenness} of $u$ is the
proportion of shortest paths that transit through an $u$ as an intermediate node.
The
\emph{eigencentrality} of $u$ is its corresponding coordinate in the largest eigenvector of the weighted adjacency matrix.  PageRank centrality is based on the
stationary distribution of a
random walk on the network that periodically teleports to a node chosen uniformly at random.

\subsection{Model Selection}\label{selec}

The goal of our model selection experiments is to determine a random graph model that matches empirically observed properties of character networks. Our methodology is to create a compact summary of
the network statistics that is invariant to the labeling of the nodes of the network. In other words, we would derive the same statistics if we permuted the adjacency matrix. The summaries we use are
the $3$-profile, $4$-profile, and eigenvalue histogram. The $k$-\emph{profile} of a graph $G$ counts the number of times each graph on $k$ nodes appears as an induced subgraph of $G$; see
\cite{elen3prof}. An \emph{eigenvalue histogram} is a histogram of the eigenvalues of the normalized Laplacian matrix, which all lie between 0 and 2, with equally spaced bins. These techniques are
well established in model selection for various types of biological and social networks~\cite{bon2,janssen}.

In contrast to the previous section we use undirected, unweighted graphs for this experiment. This choice reflects our goal to model the connectivity of the networks, rather than their joint
connectivity and weight structure.

We use the algorithm in \cite{elen4prof} to compute a global graph $4$-profile for each character network. This is a generalization of graphlets \cite{przuljGraphlets,shervashidzeGraphlets}, a
similar method of motif counting for connected subgraphs. One difference is that the 4-profile includes disconnected graphs as well. We use standard algorithms for computing all eigenvalues of the
normalized Laplacian where we treat the normalized Laplacian as a dense matrix. We compute a histogram based on five equally spaced bins.

We examine the following random graph models on $n$ nodes, with parameters chosen to match those of the original character network:
\begin{enumerate}
\item \textit{Preferential Attachment (PA)}. In the PA model, at each step, a node is added to the graph and $m$ edges are placed from the new node to existing nodes. These edges are chosen with
    probability proportional to the degree of each node before the new node arrived. If $m$ is chosen such that
\begin{align*}
\frac{2}{n} + 2m  = \frac{2|E|}{n},
\end{align*} then the number of edges will match that of the original graph in expectation.
\item The \textit{Binomial Random Graph} $G(n,p)$, or \textit{\Erdos\!-\Renyi (ER)} model. Each of the $\binom{n}{2}$ edges is connected according to an independent binomial random variable with
    probability $p$ proportional to the expected average degree. We use $p = |E|/\binom{n}{2}$ to match the average degree of the original network.
\item The \textit{Chung-Lu (CL)} model. The CL model generalizes the binomial random graph model to non-uniform edge probabilities. Graphs in this model are parameterized by an expected degree
    distribution (the character network's true degree distribution) rather than a scalar average degree. Each edge is connected with probability proportional to the product of the expected degrees
    $w_i$ of its endpoints:
\begin{align*}
p_{ij} = \frac{1}{C} w_i w_j .
\end{align*}
\item The \textit{Configuration Model (CFG)}. In the CFG model, we select a graph uniformly from the set of graphs which exactly match the target degree distribution. In practice, the degree
    distribution may vary slightly from the target since we disregard self loops and multi-edges created during this process.
\end{enumerate}

Our method to determine the best random graph model fitting the data is to generate samples and train a machine learning algorithm to identify each model. We then ask the algorithm to classify the
real graph. First, $100$ random graphs from each model are used to train a machine learning classifier. Then in the test step, the classifier predicts a class label for the original character
network. This provides a measure of which random graph model best fits the character network. We study the following machine learning algorithms: two variants of linear classifiers (SVMs) and two
ensemble methods based on decision trees (Random Forests and Boosted Decision Trees). For more about these models, see \cite{statisticalLearning}.
\begin{enumerate}
\item \textit{Support Vector Machines (SVM)}. The SVM algorithm is a simple way to classify points in Euclidean space. Geometrically, the binary SVM classifier is defined by a hyperplane $\vecw$
    that maximally separates points from both classes on either side. This problem can be formulated as a quadratic program with either $\ell_1$ or $\ell_2$ regularization.
Since our application involves more than two classes, a ``one-versus-the-rest'' classifier is trained for each random graph model. Then we select the model corresponding to the highest confidence
score during classification.
\item \textit{Random Forest}. In this algorithm, classifiers combine many weaker decision trees,  each working on a random subset of the feature space, to reduce variance and increase robustness. The output is simply a sum of the scores given by each tree. 
\item \textit{Boosted Decision Trees}. This algorithm gives another approach to combine several weak learners. We use a popular boosting algorithm called AdaBoost \cite{adaboost} in which new trees
    are constructed sequentially to correct mistakes made by the previous trees. As before, the final prediction is decided by summing across trees.
\end{enumerate}

\subsection{Data}

\paragraph{Novels:} Our method for extracting character networks from novels begins with the tokenization of an input of text. Character names and aliases are then gathered by the parser, coupled with manual
addition and subtraction as needed. Names and aliases representing one character are assigned to its main name. The main names represent the nodes in the network. The parser runs through the text
recording the occurrence of two names within a certain number of words apart. For our results, we set the distance parameter to 15 words apart. In the instance where two names share the same keystone
and are both within the specified distance along with another name, the parser will record one occurrence between the unique key names. The number of occurrences between two key names represents the
weighted edge between the corresponding nodes in the character graph. The node and adjacency lists are recorded via two separate CSV files, which are imported to Gephi, an open source software
platform for network analysis and visualization.

The following books were selected for the experiment: \textit{Twilight} by Stephanie Meyer, \textit{Harry Potter and the Goblet of Fire} by J.K.\ Rowlings, and \textit{The Stand} by Stephen King. We
summarize basic network statistics for the novels in Table~\ref{tab:metrics}. The results support the view of character networks as complex networks that are dense and small world.
\begin{table}[h]
\caption{Global metrics of character networks from the novels.}
\label{tab:metrics}
	\begin{center}
      \begin{tabular}{| r | r | r | r | r | r | r | r |} 
    \hline
    Novel & \# Nodes & Avg. Degree & Avg. Weighted Degree & Diameter & Edge Density & Avg. Distance & Clust. Coeff.  \\ \hline  
    \emph{The Stand} &39 & 14.36 & 335.33 & 3 & 0.378 & 1.66 & 0.718   \\ \hline
    \emph{Goblet} &62 & 18.55 & 305.29 & 2 & 0.304 & 1.69 & 0.746 \\ \hline
    \emph{Twilight} &27 & 9.11 & 76.37 & 4 & 0.35 & 1.74 & 0.783       \\ \hline
    \end{tabular}
\end{center}
\end{table}

\begin{figure}[h!]
\begin{center}
\includegraphics[scale=0.7]{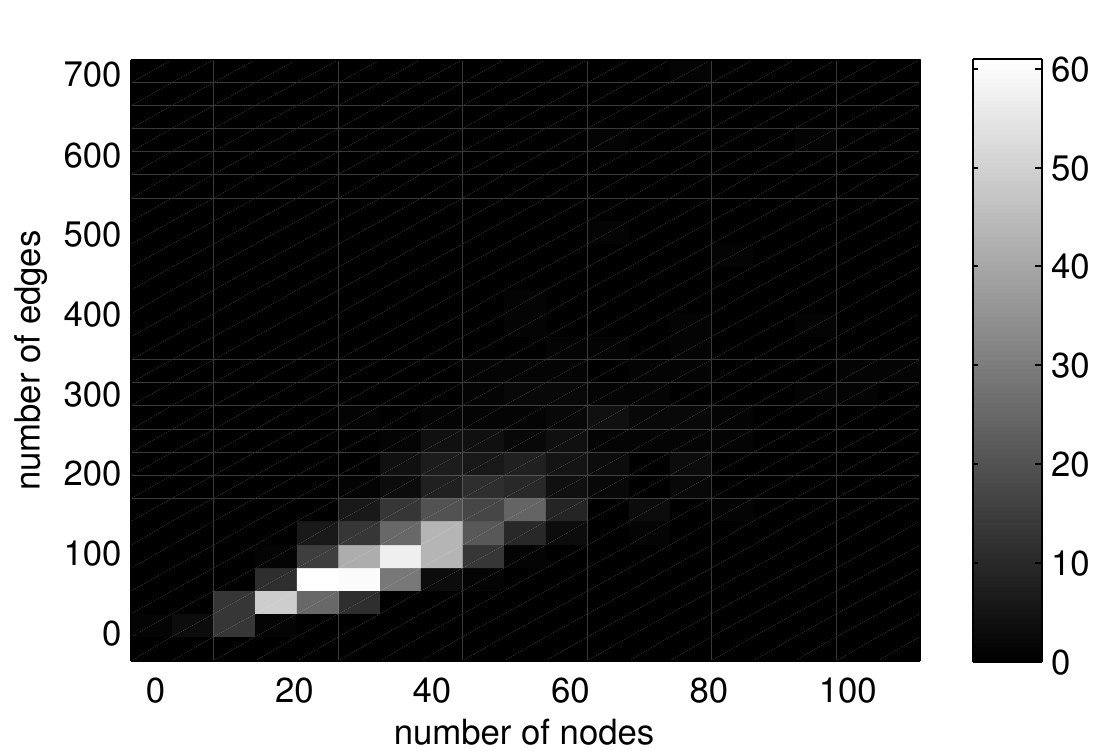}
\caption{Number of characters versus number of edges in the \emph{Moviegalaxies} network data.  The color shows how many graphs (according to the color bar) lie at the same (nodes, edges) bin.}
\label{fig:moviegalaxies-sizes}
\end{center}
\end{figure}

\paragraph{Moviegalaxies:} The website \url{http://moviegalaxies.com/} has assembled a large number of character networks based on movie scripts.
There are over 800 networks available. Each network is weighted, although we discard the weights as we only use this for the model selection problem. Some of the properties of these networks are
shown in Figure~\ref{fig:moviegalaxies-sizes}.

\section{Results}\label{results}

\subsection{Analysis of Novel Character Networks}

Main characters from each of the novels analyzed scored consistently high in each of the six centrality measures. We present the centrality measures for the top twelve characters from the novel
character networks in the figures below. Characters are ranked by increasing PageRank. For example, Harry, Ron and Hermione are identified as the top characters in \emph{Harry Potter and the Goblet
of Fire}. Further, our methods accurately predict the community structure for each of the three novels. Visualizations of the character networks and their community in the novels is found below.

\begin{figure}[h!]
	\centering
	\includegraphics[scale=0.7]{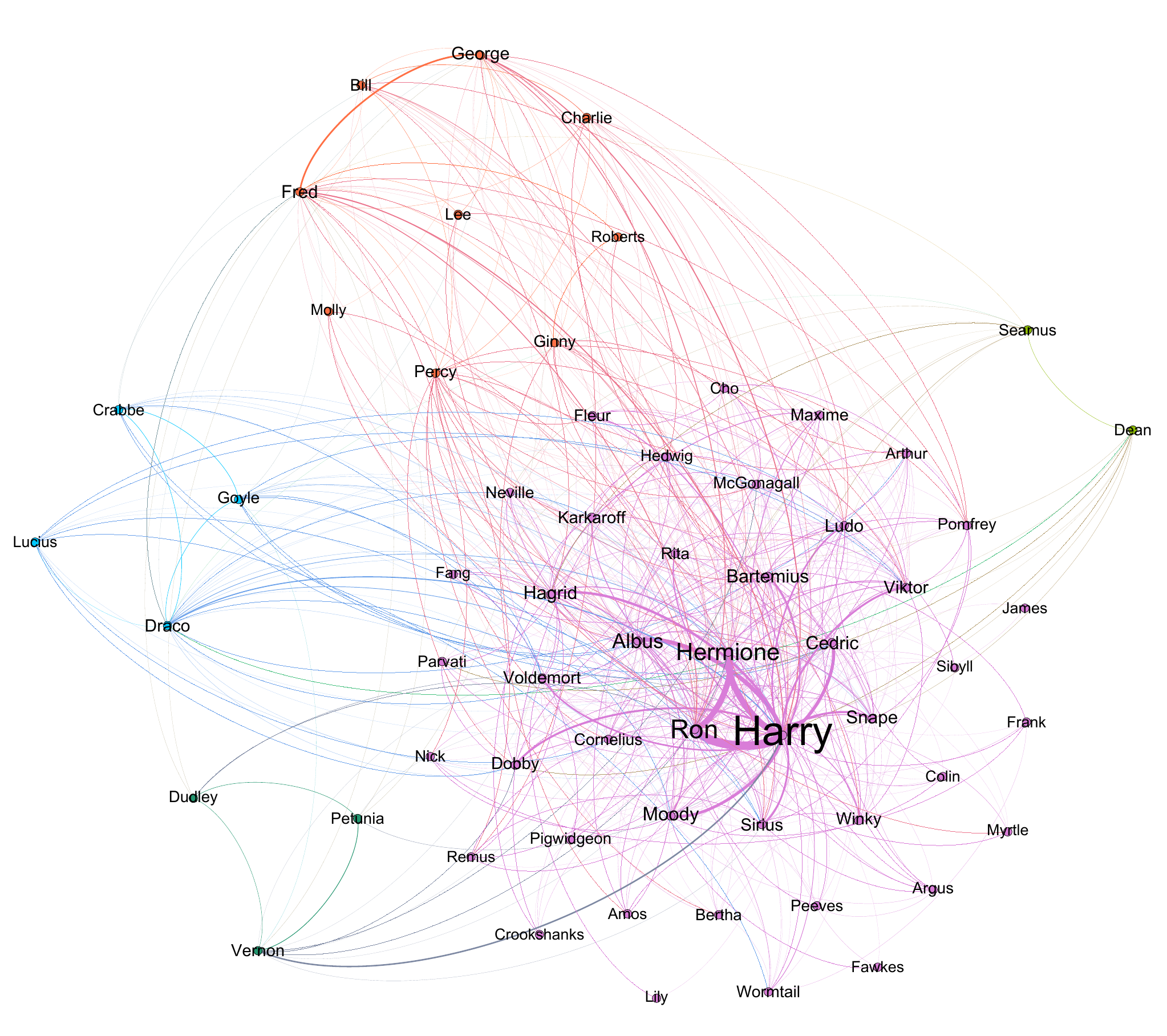}
	\caption{The character network for \textit{Harry Potter and the Goblet of Fire}. Each community is represented by a distinct color. The thickness of an edge is scaled to its weight,
and the size of a name is scaled to the Pagerank score.}\label{harryg}
\end{figure}

For \emph{Harry Potter and the Goblet of Fire} the communities were: Hogwarts, the Dursleys, the Weasleys, Sytherin, and the inseparable friends Seamus and Dean. See Figure~\ref{harryg}.
\begin{figure}[ht!]
	\begin{center}
	\includegraphics[scale=0.55]{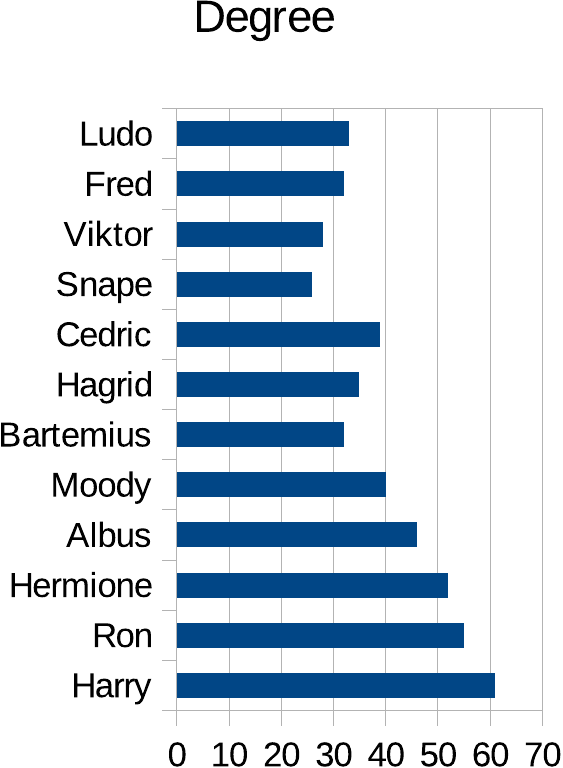}
	\includegraphics[scale=0.55]{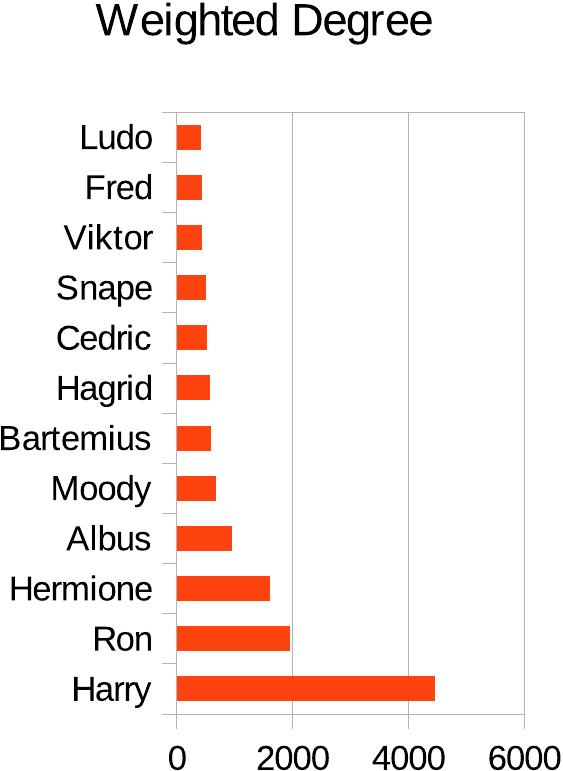}
	\includegraphics[scale=0.55]{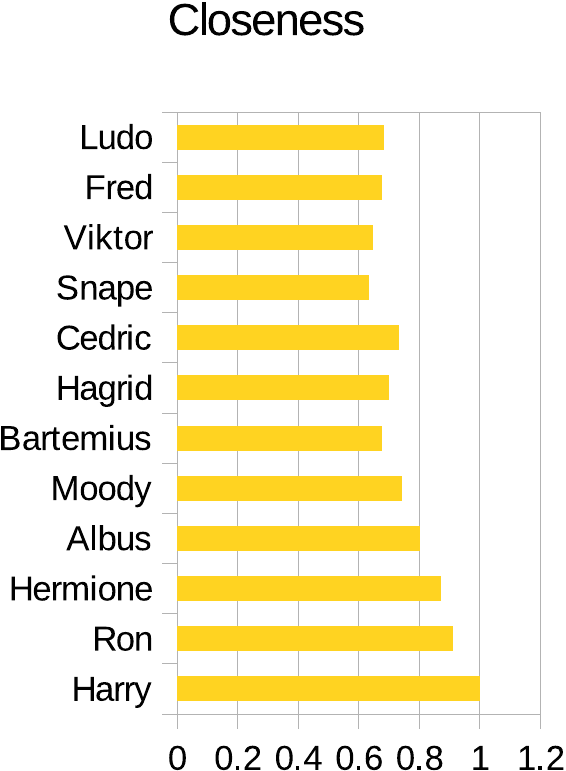}\\
	\includegraphics[scale=0.55]{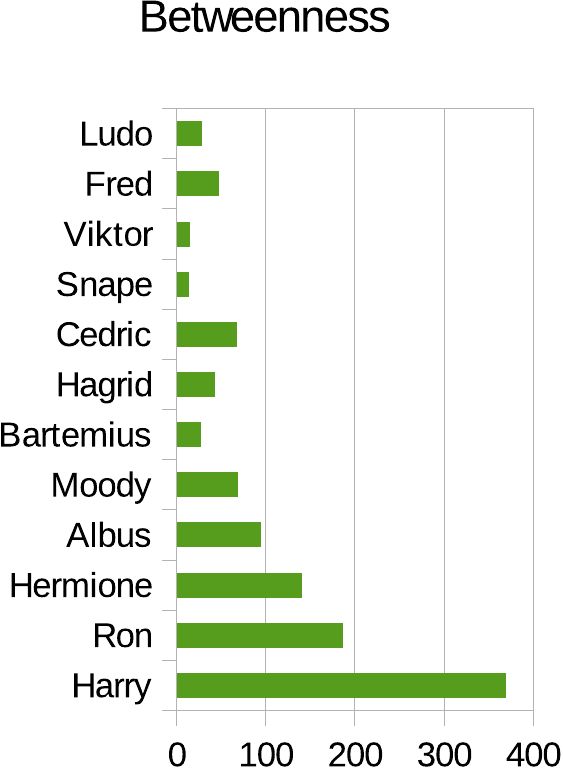}
	\includegraphics[scale=0.55]{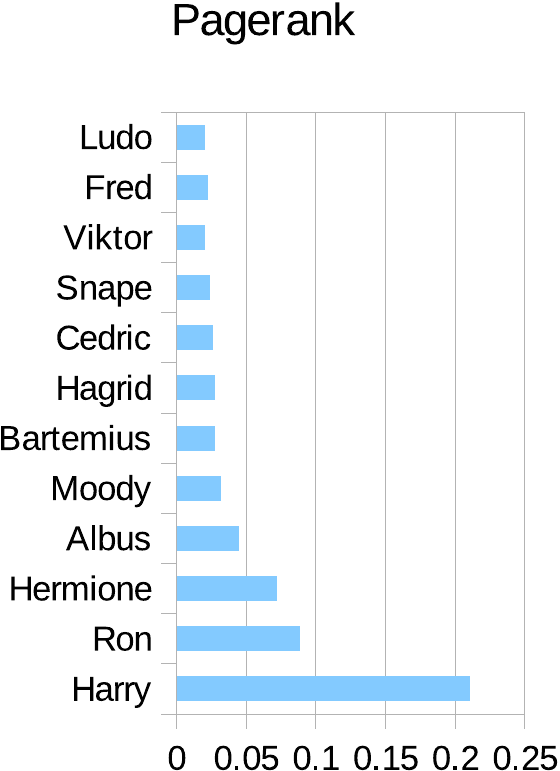}
	\includegraphics[scale=0.55]{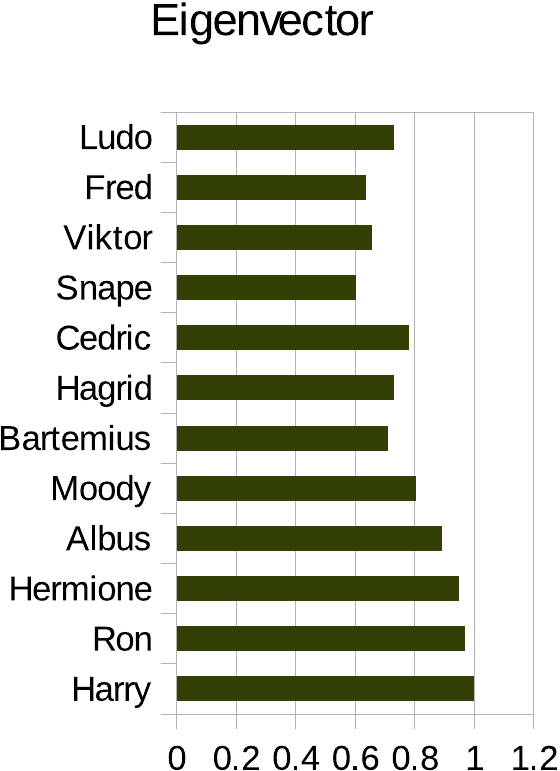}
	\caption{Centrality measures for \textit{Harry Potter and the Goblet of Fire}.}
\end{center}
\end{figure}

\newpage

\begin{figure}[h!]
	\centering
	\includegraphics[scale=0.5]{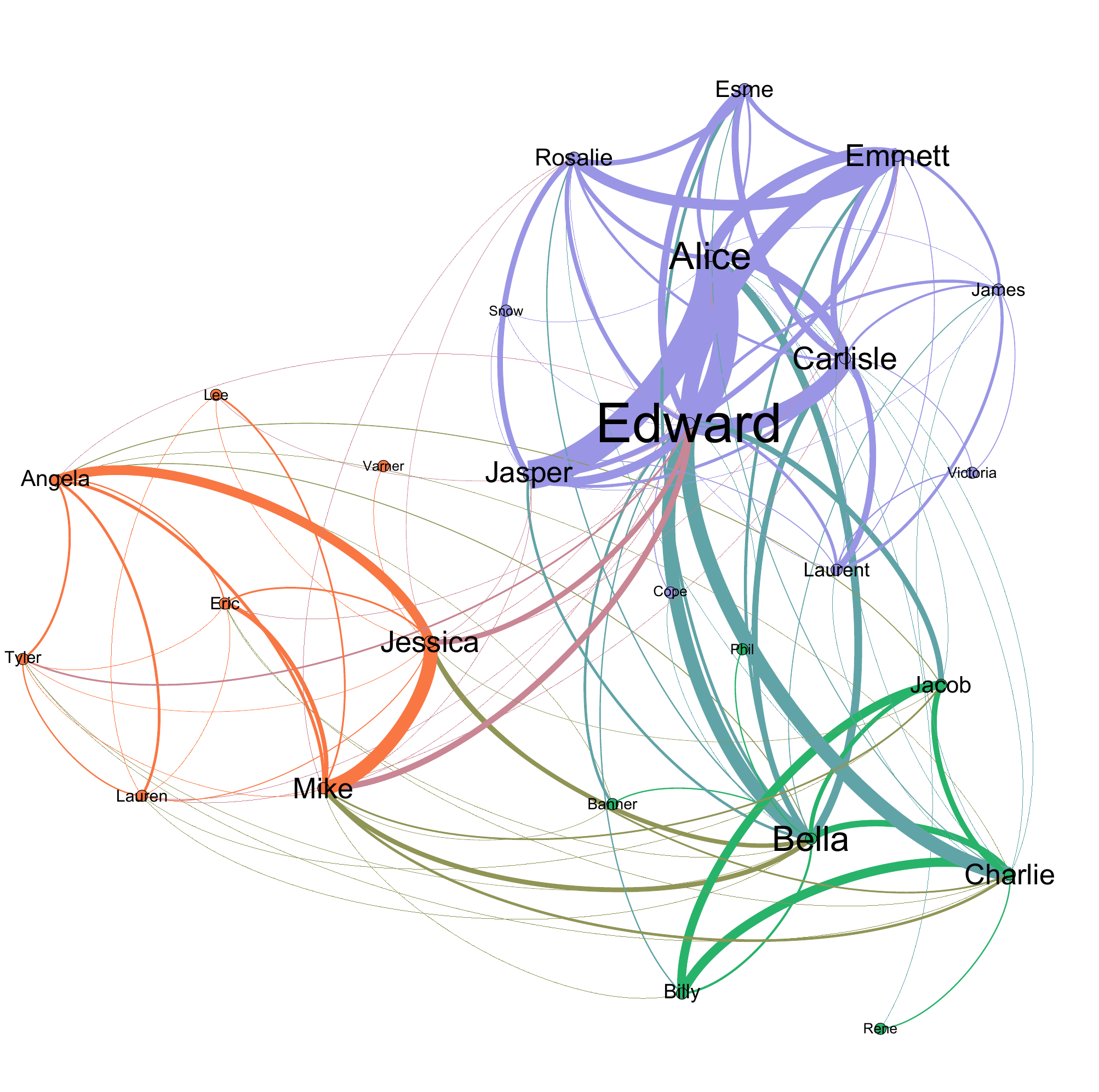}
	\caption{The character network for \textit{Twilight}.}\label{twig}
\end{figure}

For \emph{Twilight}, the three communities can be labeled as: vampires, high school students, and characters close to Charlie. See Figure~\ref{twig}.
\begin{figure}[h!]
\begin{center}
	\includegraphics[scale=0.55]{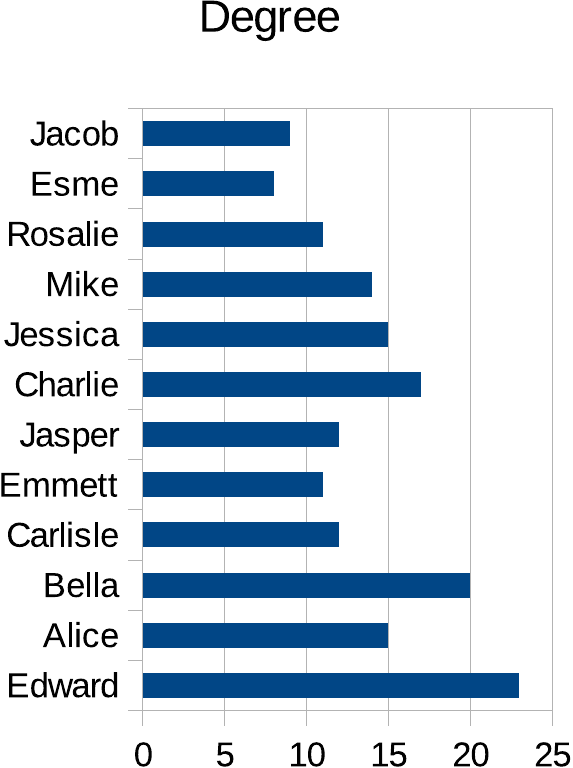}
	\includegraphics[scale=0.55]{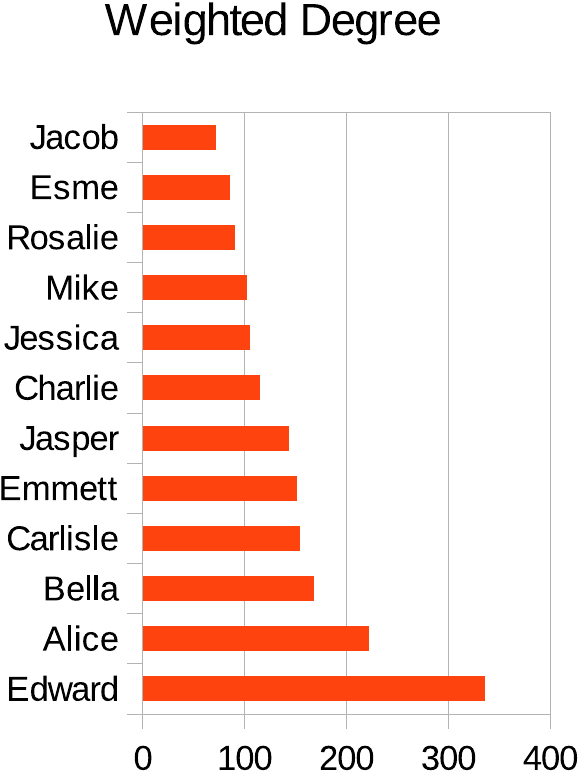}
	\includegraphics[scale=0.55]{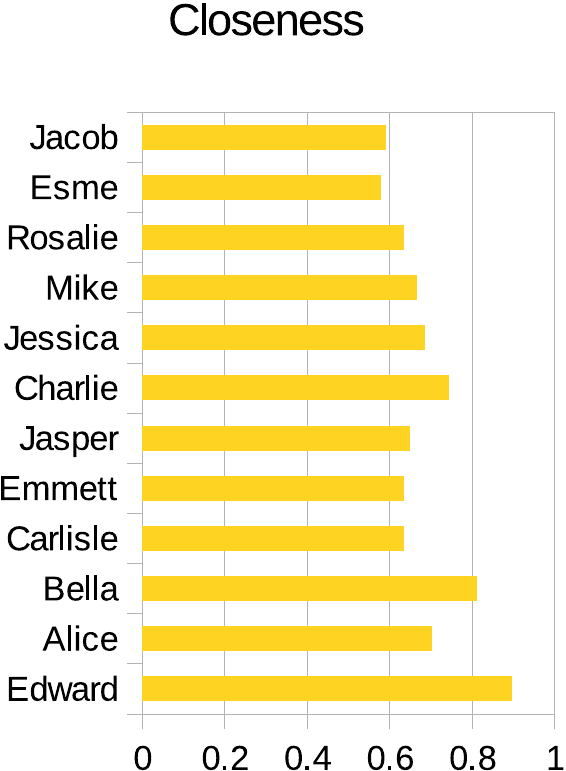}\\
	\includegraphics[scale=0.55]{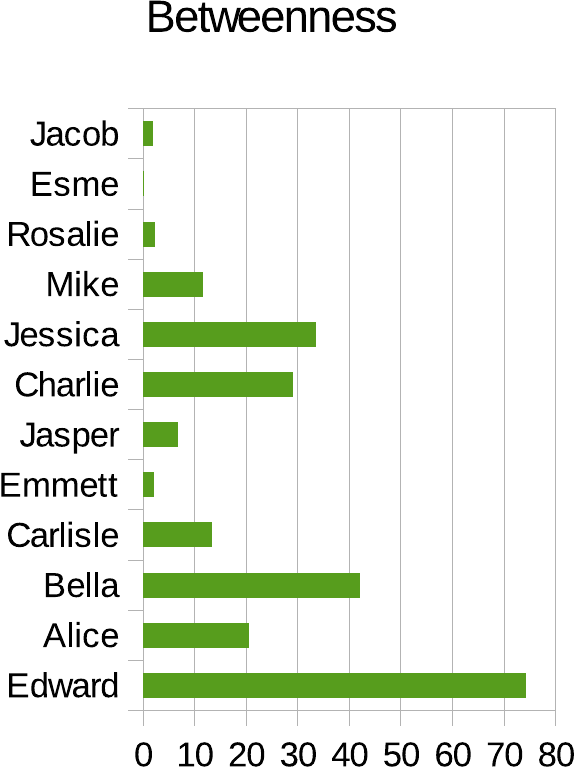}
	\includegraphics[scale=0.55]{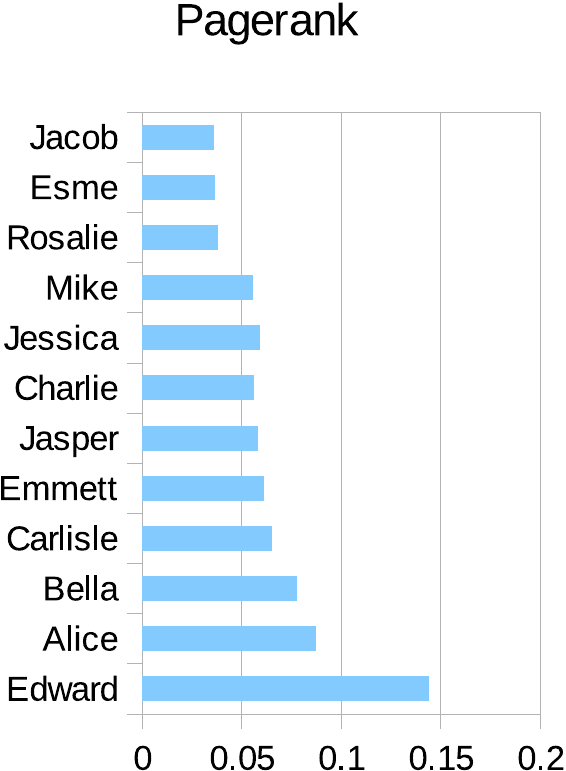}
	\includegraphics[scale=0.55]{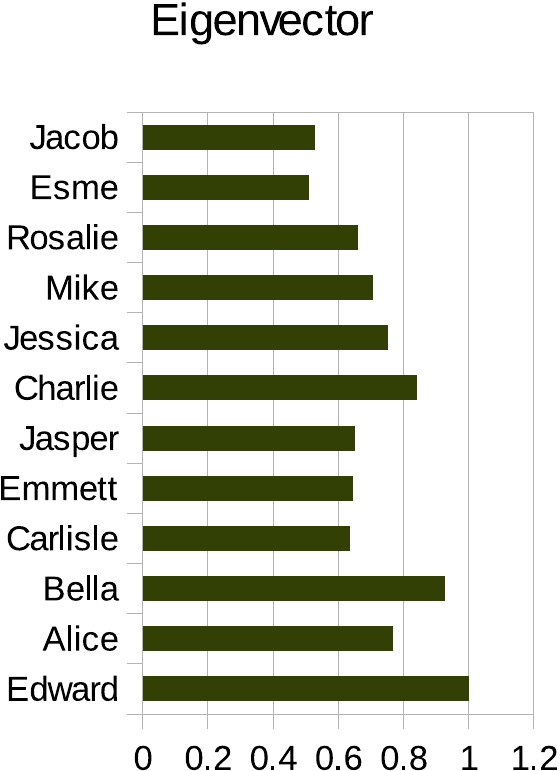}
	\caption{Centrality measures for \textit{Twilight}.}
\end{center}
\end{figure}

\bigskip

\begin{figure}[h!]
	\centering
	\includegraphics[scale=0.7]{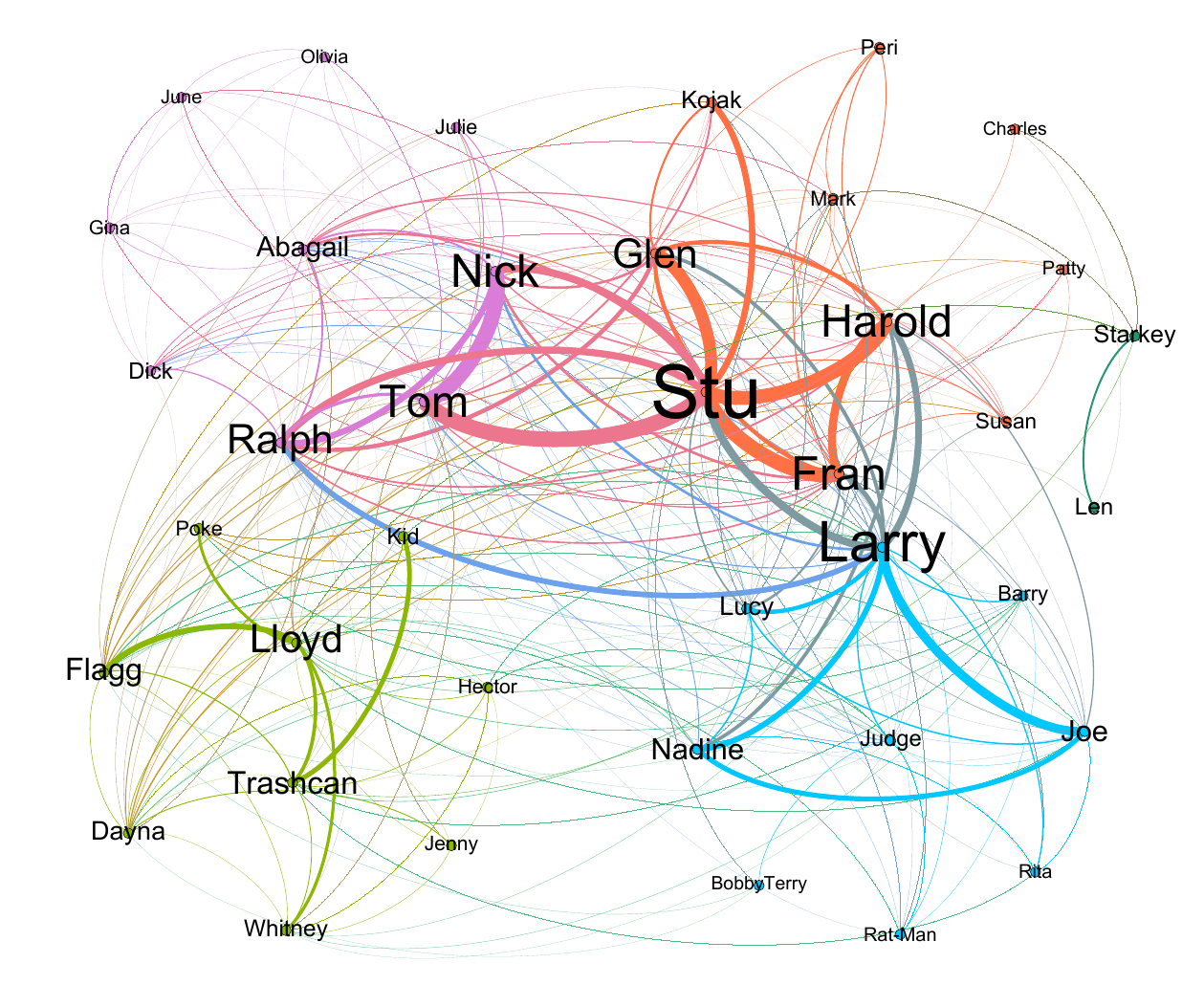}
	\caption{The character network for \textit{The Stand}.}\label{standg}
\end{figure}

For \emph{The Stand}, the government and the evil Las Vegas group emerged as separate communities. The \emph{free zone society} was divided into three groups based on their relation to the main
characters, Stu, Larry, and Nick. See Figure~\ref{standg}.
\begin{figure}[h!]
	\begin{center}
	\includegraphics[scale=0.55]{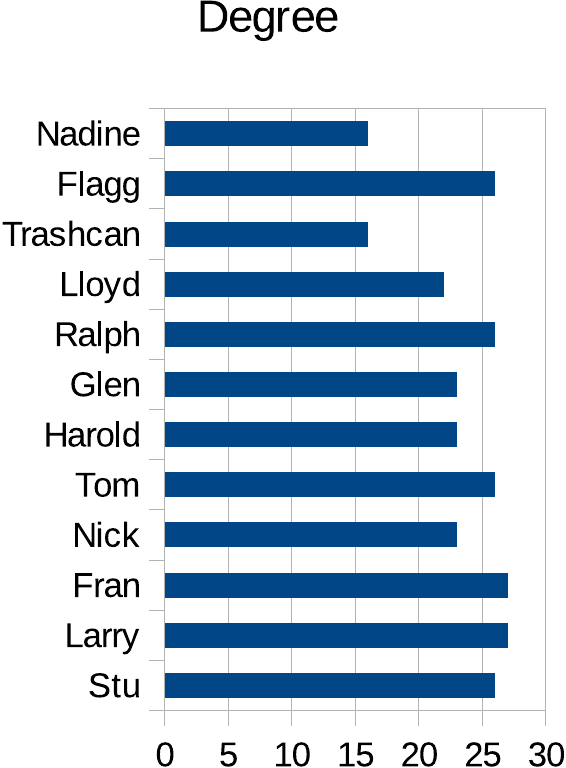}
	\includegraphics[scale=0.55]{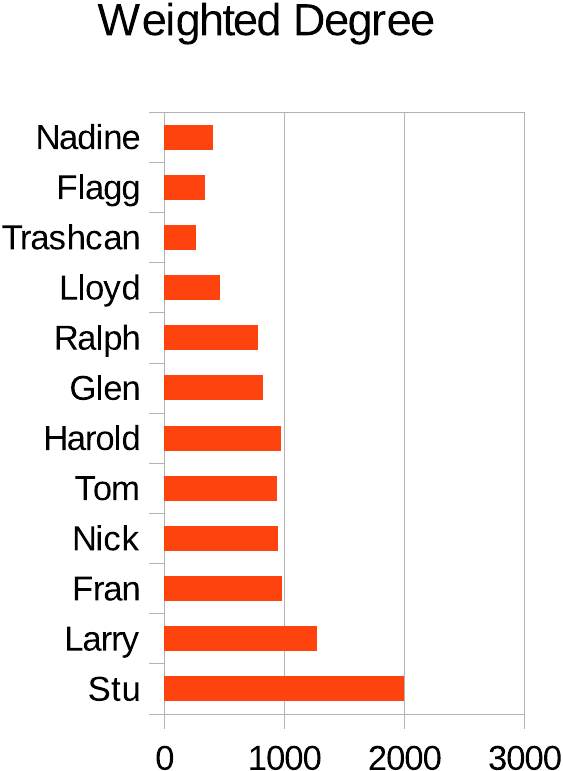}
	\includegraphics[scale=0.55]{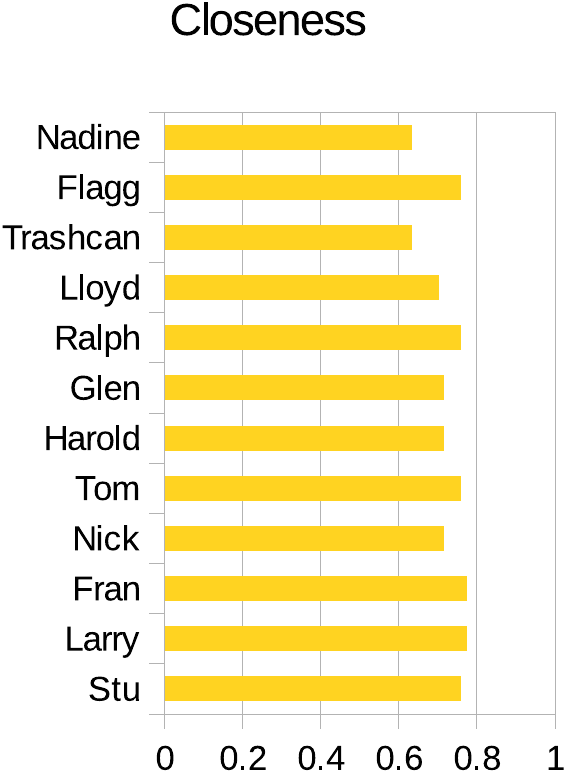}\\
	\includegraphics[scale=0.55]{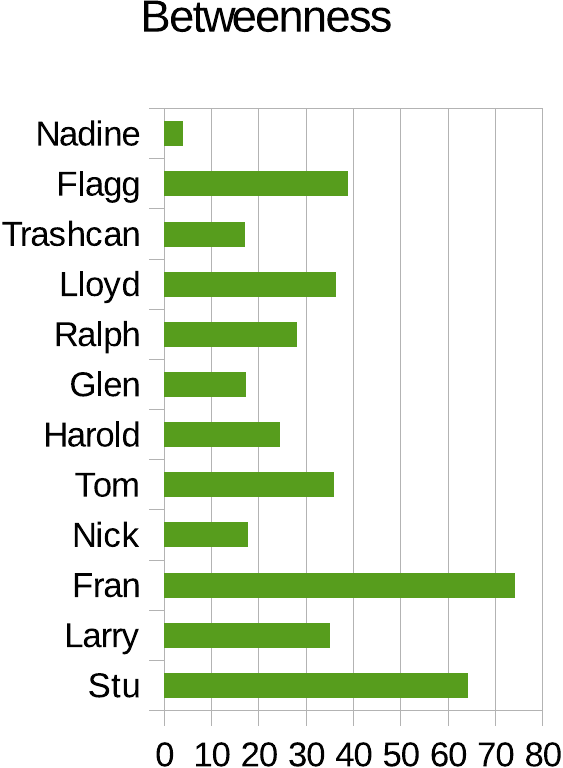}
	\includegraphics[scale=0.55]{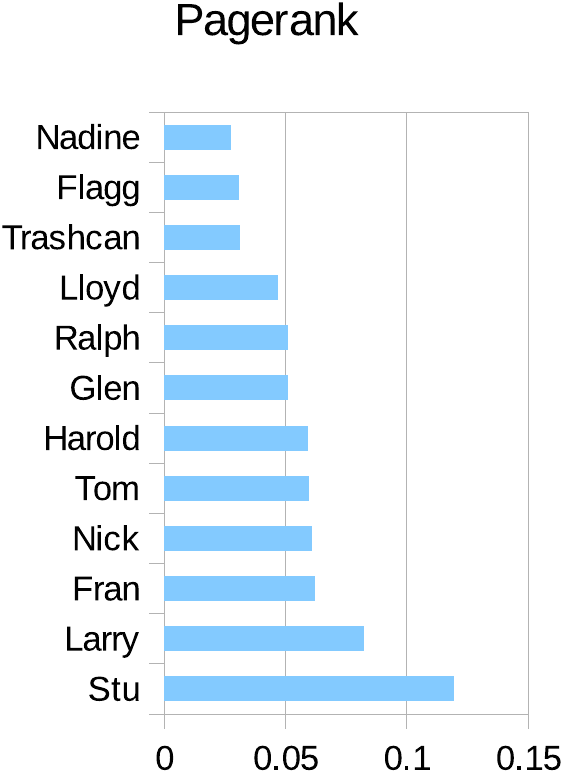}
	\includegraphics[scale=0.55]{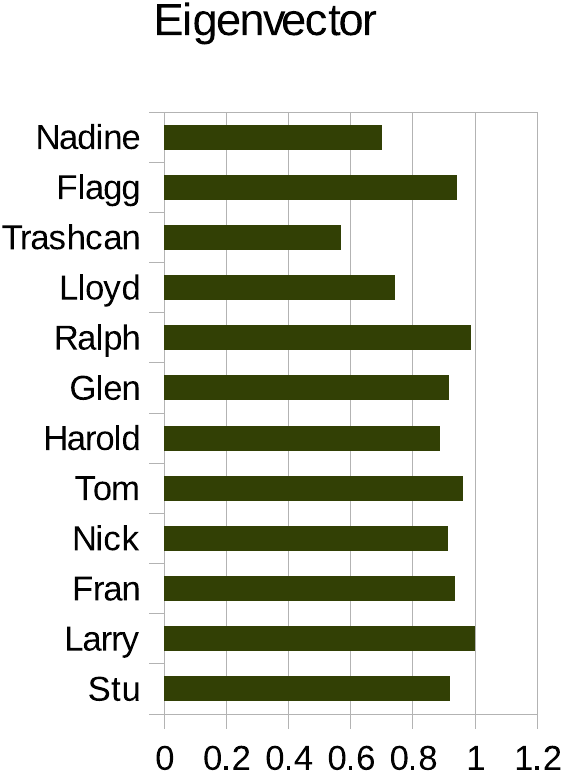}
	\caption{Centrality measures for \textit{The Stand}.}
\end{center}
\end{figure}

\subsection{Model Selection Results}
Hyperparameters for each classifier were selected using stratified, $5$-fold cross validation. All features were normalized to have zero mean and unit variance before training. First, the random
graph data was split into half training and half holdout. Classification performance on the holdout set verified both the choice of hyperparameters and the separability of classes in our chosen
feature space. All classifiers achieved nearly perfect classification on the holdout set, with over $0.98$ precision and recall in nearly all cases (often exactly $1$). The F1 score was at least
$0.97$. Thus, our four random graph models represent distinct classes.

\begin{figure}[h!]
	\centering
	\includegraphics[scale=0.55]{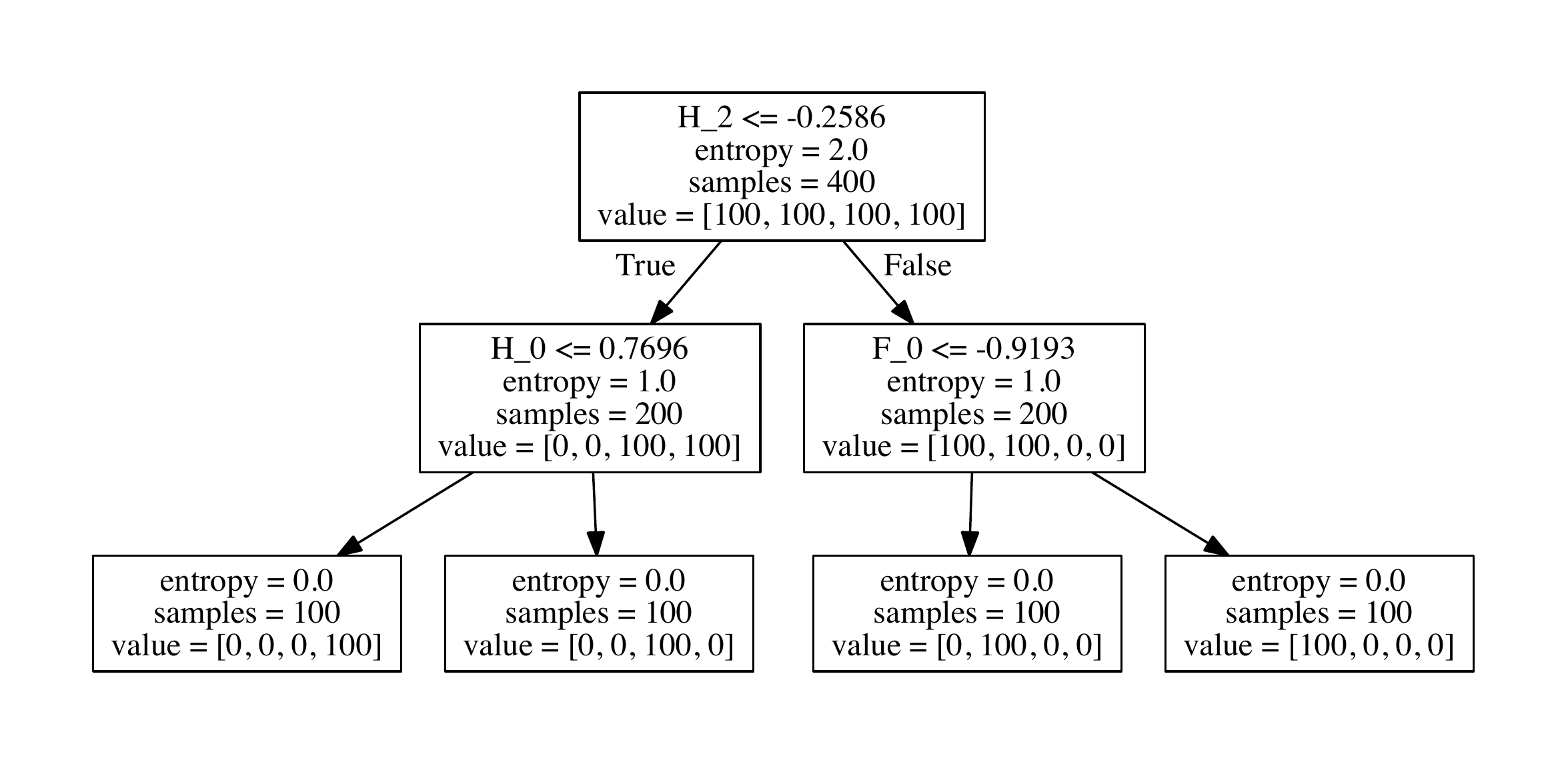}
	\caption{Example decision tree for the \emph{Goblet} graph.}\label{fig:tree}
\end{figure}

Table~\ref{tab:modelScores} shows model selection scores for the setup described in Section~\ref{selec} (we train on the entire random graph data, and test on the original character network). These
scores were calculated differently depending on the classifier. For the SVM algorithms, distance to the separating hyperplane was used. For AdaBoost, we use the final decision function, and soft
decision probabilities were used for the random forests; see Figure~\ref{fig:tree}. For all classifiers, a more positive (negative) score indicates more confidence the original graph does (not)
belong to the model. Clearly, CL is the best random graph model for all three novels, with each remaining model taking a distant second place in at least one case.

\begin{figure}[h!]
	\centering
	\includegraphics[scale=0.8]{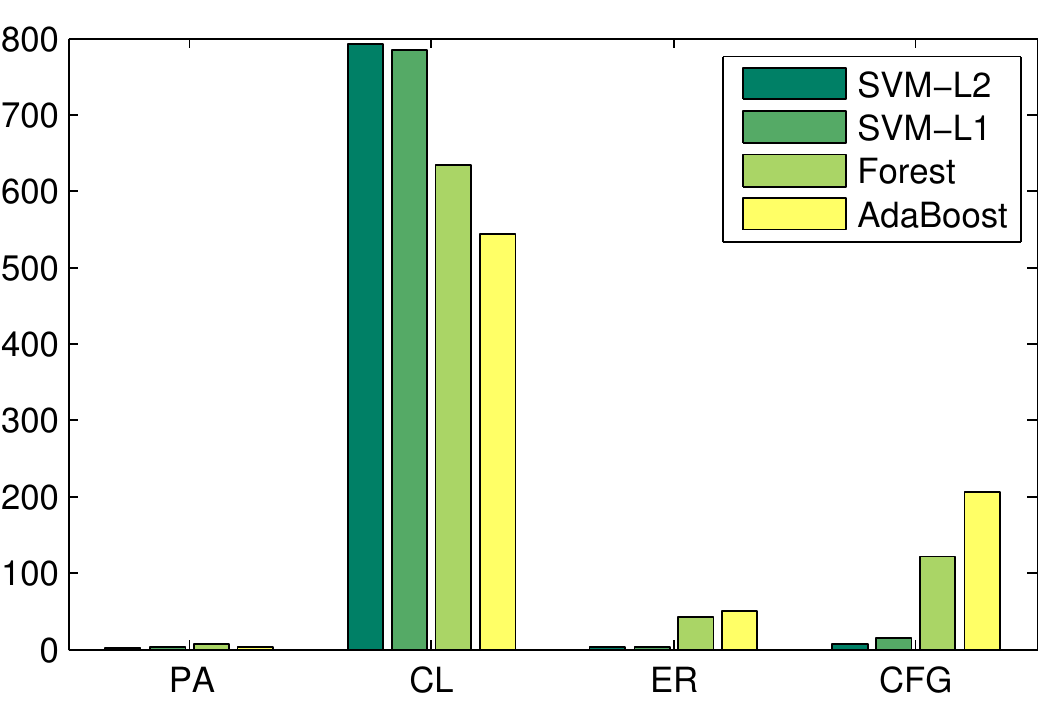}
	\caption{Summary of \emph{Moviegalaxies} model selection using graph profiles and eigenvalue histograms as features.}\label{fig:mgbar}
\end{figure}

Figure~\ref{fig:profiles} shows our naming convention for the motifs used in our graph profile features. The most important features for the CL SVM hyperplanes were predominantly cliques: induced
subgraphs $H_3$, $F_5$, $F_9$, and $F_{10}$. For the tree-based classifiers, the most important motifs for distinguishing among graph models include some disconnected subgraphs: $H_0$, $H_2$, $F_2$,
$F_{5}$, and $F_{10}$. The eigenvalue histograms generally had low importance for all machine learning classifiers. Thus, similar results were obtained using only graph profile features. See
Table~\ref{tab:modelScoresProfiles}. Figure~\ref{fig:mgbar} shows similar aggregate results for the 800 character networks in the \emph{Moviegalaxies} data set, with CL as the best random graph model
for the overwhelming number of character networks.

\begin{table}[h!]
\caption{Model selection scores for random graph models using graph profiles and eigenvalue histograms as features. CL is selected by all machine learning classifiers as the best model.}
\label{tab:modelScores}
\begin{center}
\begin{tabular}{c @{\quad} r @{\qquad} r  @{\quad} r @{\quad} r @{\quad} r}
\hline
\multicolumn{1}{c}{Novel} & \multicolumn{1}{l}{Classifier} & \multicolumn{1}{c}{PA} & \multicolumn{1}{c}{CL}& \multicolumn{1}{c}{ER}& \multicolumn{1}{r}{CFG}\\
\hline
\multirow{4}{*}{\emph{Goblet}} & SVM-$\ell_2$ 	 &  $2.78$	&   $\mathbf{4.59}$	&   $-1.10$	&   $-10.65$\\
& SVM-$\ell_1$  	&   $-0.66$	&   $\mathbf{3.81}$	&   $-1.55$	&   $-10.80$\\
& Forest  &   $0.00$	&   $\mathbf{0.91}$	&   $0.094$	&   $0.0011$\\
& AdaBoost  & $-47.2$	&   $\mathbf{47.4}$	&   $25.5$	&   $-25.7$\\
\hline
\multirow{4}{*}{\emph{Twilight}} & SVM-$\ell_2$ 	 &  $-0.671$	&   $\mathbf{4.49}$	&   $-2.98$	&   $-9.39$\\
& SVM-$\ell_1$  	&   $-3.08$	&   $\mathbf{5.19}$	&   $-2.06$	&   $-12.21$\\
& Forest  &   $0.00083$	&   $\mathbf{0.800}$	&   $0.0248$	&   $0.175$\\
& AdaBoost  & $-43.06$	&   $\mathbf{32.30}$	&   $10.74$	&   $0.0205$\\
\hline
\multirow{4}{*}{\emph{The Stand}} & SVM-$\ell_2$ 	 &  $-1.52$	&   $\mathbf{2.65}$	&   $-1.24$	&   $-3.87$\\
& SVM-$\ell_1$  	&   $-2.32$	&   $\mathbf{2.87}$	&   $-1.14$	&   $-4.97$\\
& Forest  &   $0.00$	&   $\mathbf{0.946}$	&   $0.00$	&   $0.0544$\\
& AdaBoost  & $-47.04$	&   $\mathbf{50.03}$	&   $37.83$	&   $-40.82$\\
\hline
\end{tabular}
\end{center}
\end{table}

\begin{table}[h!]
\caption{Model selection scores for random graph models using graph profiles alone as features. Once again, CL is selected by all machine learning classifiers as the best model.}
\label{tab:modelScoresProfiles}
\begin{center}
\begin{tabular}{c @{\quad} r @{\qquad} r  @{\quad} r @{\quad} r @{\quad} r}
\hline
\multicolumn{1}{c}{Novel} & \multicolumn{1}{l}{Classifier} & \multicolumn{1}{c}{PA} & \multicolumn{1}{c}{CL}& \multicolumn{1}{c}{ER}& \multicolumn{1}{r}{CFG}\\
\hline
\multirow{4}{*}{\emph{Goblet}} & SVM-$\ell_2$ 	 &  $3.18$	&   $\mathbf{4.44}$	&   $-1.15$	&   $-10.64$\\
& SVM-$\ell_1$  	&   $-0.68$	&   $\mathbf{3.81}$	&   $-1.53$	&   $-10.81$\\
& Forest  &   $0.000$	&   $\mathbf{0.998}$	&   $0.002$	&   $0.000$\\
& AdaBoost  & $-47.2$	&   $\mathbf{47.4}$	&   $25.5$	&   $-25.7$\\
\hline
\multirow{4}{*}{\emph{Twilight}} & SVM-$\ell_2$ 	 &  $-0.54$	&   $\mathbf{5.51}$	&   $-2.73$	&   $-9.52$\\
& SVM-$\ell_1$  	&   $-2.78$	&   $\mathbf{5.25}$	&   $-2.02$	&   $-12.24$\\
& Forest  &   $0.00$	&   $\mathbf{1.00}$	&   $0.00$	&   $0.00$\\
& AdaBoost  & $-39.72$	&   $\mathbf{34.51}$	&   $-7.44$	&   $12.66$\\
\hline
\multirow{4}{*}{\emph{The Stand}} & SVM-$\ell_2$ 	 &  $-1.18$	&   $\mathbf{2.58}$	&   $-1.33$	&   $-4.02$\\
& SVM-$\ell_1$  	&   $-2.35$	&   $\mathbf{2.86}$	&   $-1.14$	&   $-4.99$\\
& Forest  &   $0.00$	&   $\mathbf{0.94}$	&   $0.00$	&   $0.06$\\
& AdaBoost  & $-46.49$	&   $\mathbf{50.32}$	&   $38.36$	&   $-42.19$\\
\hline
\end{tabular}
\end{center}
\end{table}

\def\blockdist{1}
\def\edgedist{4.5}
\def\smalldist{.1}

\begin{figure}
\centering
	\subfloat[]{
    \begin{tikzpicture}[
      inner/.style={circle,draw,fill=black,inner sep=1pt, minimum size=0.72em},
      outer/.style= {draw
     }
    ]
    \matrix (w0) [matrix of nodes, outer, nodes={inner}, label=$H_0$,ampersand replacement=\&]{
      \&{} \\
      {} \& \& {}\\
    };
    \matrix (w1) [matrix of nodes, outer, nodes={inner},right=\smalldist of w0, label=$H_1$,ampersand replacement=\&]{
        \&{} \\
       {} \& \& {}\\
     };
     \matrix (w2) [matrix of nodes, outer, nodes={inner},right=\smalldist of w1, label=$H_2$,ampersand replacement=\&]{
        \&{} \\
       {} \& \& {}\\
     };
     \matrix (w3) [matrix of nodes, outer, nodes={inner},right=\smalldist of w2, label=$H_3$,ampersand replacement=\&]{
        \&{} \\
       {} \& \& {}\\
     };
    \draw[thick] (w1-2-1)--(w1-2-3);
     \draw[thick] (w2-2-3)--(w2-2-1)--(w2-1-2);
     \draw[thick] (w3-2-1)--(w3-1-2)--(w3-2-3)--(w3-2-1);
    \end{tikzpicture} \label{fig:3subgraphs}} \\
	\subfloat[]{
 	\begin{tikzpicture}[
 	inner/.style={circle,draw,fill=black,inner sep=1pt, minimum size=.72em},
 	outer/.style= {draw
 	}
 	]
 	\matrix (w0) [matrix of nodes, outer, nodes={inner}, label=below:$F_0$, column sep=10pt, row sep=10pt,ampersand replacement=\&]{
 	     {} \&{} \\
 	    {} \& {}\\
 	  };
 	  \matrix (w1) [matrix of nodes, outer, nodes={inner},right=\smalldist of w0, label=below:$F_1$, column sep=10pt, row sep=10pt,ampersand replacement=\&]{
 	      {} \&{} \\
 	         {} \& {}\\
 	    };
 	  \matrix (w2) [matrix of nodes, outer, nodes={inner},right=\smalldist of w1, label=below:$F_2$, column sep=10pt, row sep=10pt,ampersand replacement=\&]{
 	      {} \&{} \\
 	         {} \& {}\\
 	    };
 	  \matrix (w3) [matrix of nodes, outer, nodes={inner},right=\smalldist of w2, label=below:$F_3$, column sep=10pt, row sep=10pt,ampersand replacement=\&]{
 	    {} \&{} \\
 	       {} \& {}\\
 	    };
 		\matrix (w4) [matrix of nodes, outer, nodes={inner},right=\smalldist of w3, label=below:$F_4$, column sep=10pt, row sep=10pt,ampersand replacement=\&]{
 		{} \&{} \\
 		{} \& {}\\
 		};
 		\matrix (w5) [matrix of nodes, outer, nodes={inner},right=\smalldist of w4, label=below:$F_5$, column sep=10pt, row sep=10pt,ampersand replacement=\&]{
 		{} \&{} \\
 		{} \& {}\\
 		};
 		\matrix (w6) [matrix of nodes, outer, nodes={inner},right=\smalldist of w5, label=below:$F_6$, column sep=10pt, row sep=10pt,ampersand replacement=\&]{
 		{} \&{} \\
 		{} \& {}\\
 		};
 		\matrix (w7) [matrix of nodes, outer, nodes={inner},right=\smalldist of w6, label=below:$F_7$, column sep=10pt, row sep=10pt,ampersand replacement=\&]{
 		{} \&{} \\
 		{} \& {}\\
 		};
 		\matrix (w8) [matrix of nodes, outer, nodes={inner},right=\smalldist of w7, label=below:$F_8$, column sep=10pt, row sep=10pt,ampersand replacement=\&]{
 		{} \&{} \\
 		{} \& {}\\
 		};
 		\matrix (w9) [matrix of nodes, outer, nodes={inner},right=\smalldist of w8, label=below:$F_9$, column sep=10pt, row sep=10pt,ampersand replacement=\&]{
 		{} \&{} \\
 		{} \& {}\\
 		};
 		\matrix (w10) [matrix of nodes, outer, nodes={inner},right=\smalldist of w9, label=below:$F_{10}$, column sep=10pt, row sep=10pt,ampersand replacement=\&]{
 		{} \&{} \\
 		{} \& {}\\
 		};
 	\draw[thick] (w1-2-1)--(w1-2-2);
 	\draw[thick] (w2-2-2)--(w2-2-1);
 	\draw[thick] (w2-1-2)--(w2-1-1);
 	\draw[thick] (w3-1-1)--(w3-2-1)--(w3-2-2);
 	\draw[thick] (w4-1-1)--(w4-2-1)--(w4-2-2)--(w4-1-2);
 	\draw[thick] (w5-1-1)--(w5-2-1)--(w5-2-2)--(w5-1-1);
 	\draw[thick] (w6-1-1)--(w6-2-1)--(w6-2-2);
 	\draw[thick] (w6-1-2)--(w6-2-1);
 	\draw[thick] (w7-1-1)--(w7-2-1)--(w7-2-2)--(w7-1-2)--(w7-1-1);
 	\draw[thick] (w8-1-1)--(w8-2-1)--(w8-2-2)--(w8-1-2)--(w8-2-1);
 	\draw[thick] (w9-1-1)--(w9-2-1)--(w9-2-2)--(w9-1-2)--(w9-2-1)--(w9-1-1)--(w9-1-2);
 	\draw[thick] (w10-1-1)--(w10-2-1)--(w10-2-2)--(w10-1-2)--(w10-1-1)--(w10-2-2)--(w10-1-2)--(w10-2-1);
  \end{tikzpicture} \label{fig:4subgraphs}}
\caption{(a) The four non-isomorphic graphs on $3$ nodes that comprise the graph $3$-profile. (b) The eleven non-isomorphic graphs on $4$ nodes that comprise the graph $4$-profile.}
 \label{fig:profiles}
 \end{figure}
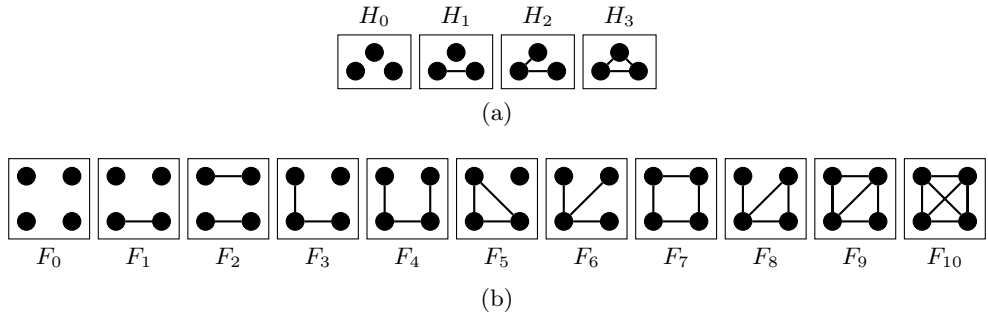

\section{Discussion and Future Work}\label{conc}

We presented a comparative and quantitative analysis of character networks arising from various novels and films. In particular, we analyzed the weighted social networks from the novels
\emph{Twilight} \emph{The Stand}, and \emph{Harry Potter and the Goblet of Fire}, along with social networks from 800 films catalogued by \cite{movieGalaxies}. For each of the character
networks from the three novels, we extracted the social network from co-occurrence of character names. Community structure was extracted for each network, and statistics such as PageRank and various
centrality measures were computed for the characters. In each case, our methodology extracts accurate literary conclusions from the data sets, and successfully identifies the influential characters
and the constellations of lesser characters in the books. As pointed out first in~\cite{beveridge}, the analysis provided of these texts was done algorithmically, without resort to conventional
literary analysis.

For both the novel and \url{http://moviegalaxies.com/} data sets, we employed machine learning techniques to compare and contrast the models against simulated data from popular complex network
models. The models considered were the Chung-Lu (CL) model, the configuration model, the PA model, and binomial random graphs. Our methodology used small subgraph counts or motifs as classifiers for
the Support Vector Machine (SVM) and other machine learning algorithms. For all the data sets, SVM and the other algorithms clearly separated the models, and indicated that the CL model provided the
best alignment with the data.

There are various explanations for the conclusions derived from the model selection experiments. As the character networks we consider have relatively few nodes, they are less likely to exhibit
various properties such as power law degree distributions or dimensionality found in various on-line social networks such as Facebook. Hence, preferential attachment (an early and successful adopted
model for complex networks) or geometric models may be less relevant for character networks. The CL model has a number of properties amenable to modeling character networks. From a literary
perspective, an author may intuit a hierarchy of character influence (separated by the degrees of the nodes representing characters), then randomly generate the social ties in the fictional work to
complete the network. For instance, Rowlings may have decided in the Harry Potter series that the main triad was Harry, Hermione and Ron, and then gradually added lesser characters revolving around
this triad. In terms of the various models, the CL model has 4-node subgraph counts that more accurately model character networks. This is likely due to the property of CL graphs that they have a
more diverse set of dense subgraph structures that are more closely related to those that appear in character networks. We plan to continue investigating this finding that CL graphs are good matches
for character networks.

In future work, we plan on expanding our analysis of literary works using Project Gutenberg and other sources. We will also explore other models such as random geometric graphs and Kronecker graphs.
More broadly, our approach and those of other recent works \cite{alice,beveridge,shapes,rings}, represents a trend towards the algorithmic and big data-theoretic analysis of cultural works. Such a
direction may lead to new models for the evolution and construction of character networks, and a broader view of such networks as complex and evolving.

\end{document}